\documentclass[prb,twocolumn,amssymb,showpacs]{revtex4}
\usepackage{graphicx,amsmath,mathrsfs}
\begin{document}

\title{Helical Dirac-Majorana interferometer in a superconductor-topological insulator sandwich structure}

\author{Chao-Xing Liu$^{1,2}$ and Bj${\rm \ddot{o}}$rn Trauzettel$^1$ }

\affiliation{ $^1$Institute for Theoretical Physics and Astrophysics,
University of W$\ddot{u}$rzburg, 97074 W$\ddot{u}$rzburg, Germany;\\
 $^2$Physikalisches Institut (EP3), University of W$\ddot{u}$rzburg, 97074 W$\ddot{u}$rzburg, Germany; }

\date\today

\begin{abstract}
In the heterostructure composed of a topological insulator sandwiched by two s-wave superconductors, the time reversal
invariant topological superconducting phase, possessing helical Majorana edge modes, is found to exist
when the two s-wave superconductors form a Josephson junction with a $\pi$ phase shift. Based on such a heterostructure,
a helical Dirac-Majorana interferometer is proposed to directly measure a unique transport signature of the helical Majorana modes. Furthermore, we envision how our proposal can be realized on the basis of existing materials such as Bi$_2$Se$_3$ or Bi$_2$Te$_3$ thin films.
\end{abstract}


\maketitle

Recently, due to the discovery of topological insulators (TIs), the search for
new topological phases has attracted a lot of interests in the condensed matter physics community\cite{hasan2010,qi2010,qi2010a,moore2010}. Interestingly, topologically non-trivial phases not only exist in insulators, but also in superconductors (SCs), which are dubbed topological superconductors (TSCs). For TIs, the unique feature is the existence of topologically protected edge states (or surface states), and similarly, there are also gapless modes
at the edge (surface) of the TSCs which are Majorana fermions due to particle-hole symmetry.
Bound Majorana modes in a vortex core and chiral Majorana modes along the edge were first
proposed in $p_x$+$ip_y$ SCs\cite{read2000} and later on helical Majorana (HM) modes have been suggested to exist in TSCs with
time reversal symmetry\cite{qi2009b,tanaka2009,sato2009}.

Majorana fermions may serve as a building block for topological quantum computation
due to their non-Abelian statistics\cite{ivanov2001,nayak2008}. However, unlike TIs, up to now, the existence
of TSCs and Majorana fermions has not been experimentally confirmed.
The difficulty of the detection of Majorana fermions lies in
the fact that they are charge neutral and have no direct response to electromagnetic fields.
Therefore, recently it has been suggested to detect Majorana fermions through a Dirac-Majorana
converter, converting a pair of Majorana fermions into one Dirac fermion and vice versa\cite{fu2009,akhmerov2009}.
Such a Dirac-Majorana converter has been first proposed for the detection of chiral Majorana modes. However, for the detection of HM modes, realistic proposals barely exist\cite{footnote1}.

In this paper, we consider the heterostructure of a TI thin film sandwiched by two SCs.
Due to the proximity effect on both the top and bottom surfaces of the TI, it is found that,
if the s-wave pairing functions at the top and bottom surfaces have a relative $\pi$ phase shift,
a topological superconducting phase emerges in the TI thin film possessing a HM mode along its edges.
Based on such a configuration, we propose an experimental setup for a HM interferometer
and discuss the transport properties which can be used to distinguish it from other types of Majorana interferometers.

\begin{figure}
    \begin{center}
        \includegraphics[width=3.5in]{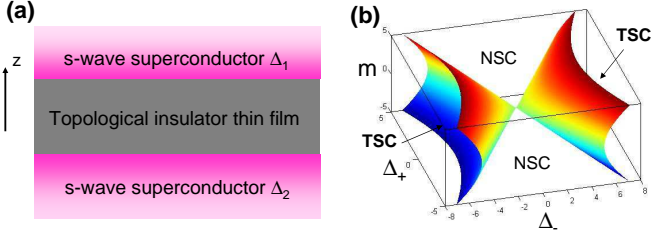}
    \end{center}
    \caption{ (Color online) (a) Side view of the SC-TI-SC heterostructure; (b) phase diagram for the SC-TI-SC heterostructure
    when the chemical potential $\mu=0$.
        }
    \label{fig:configuration}
\end{figure}

How can the envisioned heterostructure schematically shown in Fig.~\ref{fig:configuration}(a) be constructed based on existing materials?
Superconductivity can be realized by copper doped Bi$_2$Se$_3$\cite{wray2010}, TlBiTe$_2$\cite{yan2010},
and Bi$_2$Te$_3$ under pressure\cite{zhang2011}, which can be straightforwardly integrated into TI thin films based on Bi$_2$Se$_3$ or Bi$_2$Te$_3$. Let us describe the minimal model that captures the essential physics of such a system. We assume that the chemical potential lies in the bulk gap of the TI, then the low energy physics of the TI film
is described by two Dirac cones at the top and bottom surfaces, given
by the four band Hamiltonian\cite{liu2010,lu2010}
\begin{eqnarray}
	&&\hat{H}_0=\sum_k\psi^{\dag}_kh(k)\psi_k ,\nonumber\\
	&&h(k)
	=m{\tau_x}+\hbar v_f\left( k_x\sigma_y\tau_z-k_y\sigma_x\tau_z \right)
	\label{eq:hk}
\end{eqnarray}
with the field operator $\psi_k=[c_{1\uparrow},c_{1\downarrow},c_{2\uparrow},c_{2\downarrow}]^T$
, where 1(2) denotes the top (bottom) surface and $\uparrow$
($\downarrow$) denotes spin up (spin down); $v_f$ is the Fermi velocity and
$m$ the hybridization between the top and bottom surface states.
Here, the Pauli matrices $\sigma_i$ denote the spin and $\tau_i$ the opposite surfaces.
The Hamiltonian (\ref{eq:hk}) satisfies time reversal symmetry,
$T\hat{H}_0T^{-1}=\hat{H}_0$, with the time reversal operator
$T=i\sigma_yK$ ($K$ denotes complex conjugation).
The proximity effect from the s-wave SC for the top and bottom surfaces
can be taken into account by the Bogoliubov-de Gennes (BdG) Hamiltonian
\begin{eqnarray}
	&&\hat{H}_{BdG}=\sum_k
	\hat{\Psi}^\dag_kH_{BdG}(k)\hat{\Psi}_k,
	\hat{\Psi}_k=\left(
	\begin{array}{cc}
		\psi_k\\(\psi^\dag_{-k})^T
	\end{array}
	\right),\nonumber\\
        &&H_{BdG}(k)=\left(
	\begin{array}{cc}
		h(k)-\mu&\Delta\\
		\Delta^\dag&-h^T(-k)+\mu
	\end{array}
	\right)
	\label{eq:HamBdG}
\end{eqnarray}
with the s-wave pairing function $\Delta$ given by
\begin{eqnarray}
	\Delta=\left(
	\begin{array}{cc}
		i\Delta_1\sigma_y&0\\
		0&i\Delta_2\sigma_y
	\end{array}
	\right),
	\label{eq:Delta}
\end{eqnarray}
where $\Delta_{1(2)}$ is for the top (bottom) surface.
The BdG Hamiltonian is invariant under charge conjugation $C=\lambda_xK$,
where Pauli matrices $\lambda_i$ denote particle and hole.
To preserve time reversal symmetry, the pairing function $\Delta_{1(2)}$
can only be real, indicating that the relative phase of the pairing functions between
the top and bottom surfaces can only be 0 or $\pi$.
The band dispersion of the Hamiltonian (\ref{eq:HamBdG}) can be directly solved and
we first look at some simplified limits. If the chemical potential $\mu=0$,
the band dispersion is given by
\begin{eqnarray}
	E_{rts}=s\sqrt{\hbar^2v_f^2k^2+E_g^2}, E_g=\sqrt{m^2+\Delta_+^2}+t|\Delta_-|
	\label{eq:eigHamBdG}
\end{eqnarray}
with $r,t,s=\pm$ and $\Delta_\pm=\frac{\Delta_1\pm\Delta_2}{2}$.
The above band dispersion shows double degeneracy and the two degenerate
branches ($r=\pm$) can be traced back to time reversal symmetry (Kramers partners).
We note that for the branches with $t=+$, there is always a gap for any non-zero $m$, $\Delta_\pm$, while
for the branches with $t=-$, the gap will close when $m^2+\Delta_+^2=\Delta_-^2$.
At the critical point, the band dispersion (\ref{eq:eigHamBdG}) becomes linear
and there are totally two Dirac cones ($r=\pm$) which are Kramers partners.
For one Kramers partner, a single Dirac cone will induce the change of the
Berry phase by $\pi$ between the $E_g>0$ and $E_g<0$ regimes, while for the other one,
the change of the Berry phase is $-\pi$. Therefore, similarly to the quantum spin Hall
case \cite{bernevig2006d}, a topological phase transition is expected to happen across the point $E_g=0$ in parameter space.
When the SC gap $\Delta_\pm$ is much smaller than the hybridization gap $m$,
the system should be a trivial SC. Thus, we expect that the system should be a topologically non-trivial SC
when $m^2+\Delta_+^2<\Delta^2_-$ as shown in Fig. \ref{fig:configuration}(b) (which will be confirmed below).
We note that the above condition implies if $\Delta_1\Delta_2<0$. Hence, the non-trivial
phase only exists when the pairing functions of the s-wave SCs for the top and
bottom surfaces have opposite signs (i.e. they form a Josephson junction with a $\pi$ phase shift).

To determine the type of TSC, we consider the limit $\Delta_+=0$,
in which we can transform the Hamiltonian (\ref{eq:HamBdG}) into the basis
$c_{\pm,\uparrow(\downarrow)}
=\frac{1}{\sqrt{2}}\left( c_{1,\uparrow(\downarrow)}\pm c_{2,\uparrow(\downarrow)} \right)$,
and find the obtained Hamiltonian is block diagonal with the two blocks related by time
reversal symmetry. Each block of the new Hamiltonian is exactly the Hamiltonian for
the chiral TSC  discussed, for instance, in Ref.~\onlinecite{qi2010c}.
Therefore, our system is expected to be nothing but a TSC with HM edge modes. This can be seen
in analogy to the quantum spin Hall state with helical edge states consisting of two copies of the
chiral quantum anomalous Hall state\cite{bernevig2006d}.
To substantiate the above picture, we take into account another
limit ($m=0$) in which the two surfaces are decoupled from each other.
Then, for one surface, our Hamiltonian is the same as the one discussed by Fu and Kane in Ref.~\onlinecite{fu2008}.
According to Fu and Kane, a Josephson junction with a $\pi$ phase shift will induce a helical
Majorana wire along the one-dimensional junction. In this decoupled regime ($m=0$), the prediction by Fu and Kane is consistent with ours that a HM wire emerges at the edge of our sandwich structure if there is a $\pi$ phase shift across the SC-TI-SC Josephson junction.

Next, we directly solve the Hamiltonian (\ref{eq:HamBdG}) in a half infinite plane ($y<0$)
to obtain the effective Hamiltonian for the HM mode. We assume that the x-direction is translation invariant so that $k_x$ is a good quantum number.
We will first solve the eigen value problem at $k_x=0$
\begin{eqnarray}
	H_{BdG}(k_x=0,-i\partial_y)\Phi(y)=E\Phi(y)
	\label{eq:eigeneqn}
\end{eqnarray}
and then project the Hamiltonian (\ref{eq:HamBdG})
onto the subspace of the eigenstates at $k_x=0$. In order to impose open boundary conditions,
we need to add a quadratic term into the hybridization term $m=m_0+B\partial_y^2$,
where $B$ is considered to be a small positive number.
For simplicity, we only consider the case $\mu=0$, $\Delta_+=0$ and $\Delta_->m_0>0$.
With such a simplication, we search for the zero-energy state ($E=0$) with the ansatz
$\Phi\propto e^{\lambda y}\phi$ for the boundary conditions $\Phi(0)=0$ and $\Phi(y\rightarrow-\infty)=0$,
and find the doubly degenerate eigenstates
\begin{eqnarray}
	\Phi_i=\frac{1}{N_0}\left( e^{-\lambda_1y}-e^{-\lambda_2y} \right)\phi_i,\qquad i=\pm,
	\label{eq:eigenstate}
\end{eqnarray}
where $N_0$ is a normalization factor and $\lambda_{1,2}$ given by
\begin{eqnarray}
	\lambda_{1,2}=\frac{1}{2B}\left( -\hbar v_f\pm\sqrt{\hbar^2v_f^2-4B(m_0+\Delta_-)} \right)
	\label{eq:lambda}
\end{eqnarray}
which are real numbers, smaller than zero. The wave functions $\phi_{+}=\frac{1}{4}\left[1+i,1-i,-1-i,1-i,1-i,1+i,-1+i,1+i \right]^T$ and
$\phi_{-}=\frac{1}{4}\left[-1-i,1-i,-1-i,-1+i,-1+i,1+i,\right.$ $\left.-1+i,-1-i \right]^T$ are given in the basis
$|1,\uparrow,e\rangle, |1,\downarrow,e\rangle, |2,\uparrow,e\rangle, |2,\downarrow,e\rangle, |1,\uparrow,h\rangle,
|1,\downarrow,h\rangle, |2,\uparrow,h\rangle, |2,\downarrow,h\rangle$.
Since the $\phi_\pm$ are invariant under charge conjugation ($C\phi_{\pm}=\phi_\pm$)
and can be related to each other by time reversal ($T\phi_+=-\phi_-$ and $T\phi_-=\phi_+$),
they are expected to be HM modes.
Indeed, for nonzero $k_x$, we project the Hamiltonian
(\ref{eq:HamBdG}) into the subspace of $\Phi_\pm$, which can be done by
expanding the field operator $\hat{\Psi}$ as $\hat{\Psi}(k_x,y)=\sum_i\Phi_i(y)\hat{\gamma}_i(k_x)$ ($i=\pm$),
and obtain the effective Hamiltonian
\begin{eqnarray}
	\hat{H}_{\rm eff}=\sum_{k_x}\hbar v_fk_x\Bigl( \hat\gamma_+^\dag(k_x)\hat\gamma_+(k_x)-\hat\gamma_-^\dag(k_x)\hat\gamma_-(k_x) \Bigr),
	\label{eq:HMES_Heff}
\end{eqnarray}
which is exactly the HM mode as expected.

After determining the condition of the topological phase for $\mu=0$,
we can easily extend to the $\mu\neq 0$ regime. This is done by analyzing the $\mu\neq 0$ case and afterwards adiabatically connecting the result to the $\mu=0$ regime. By solving the bulk dispersion, the helical topological superconductor phase is found to exist
in the regime $|m|<\sqrt{\left( \frac{\mu^2}{\Delta^2_-}+1 \right)\left(
\Delta_-^2-\Delta_+^2\right)}$.

\begin{figure}
    \begin{center}
        \includegraphics[width=3.3in]{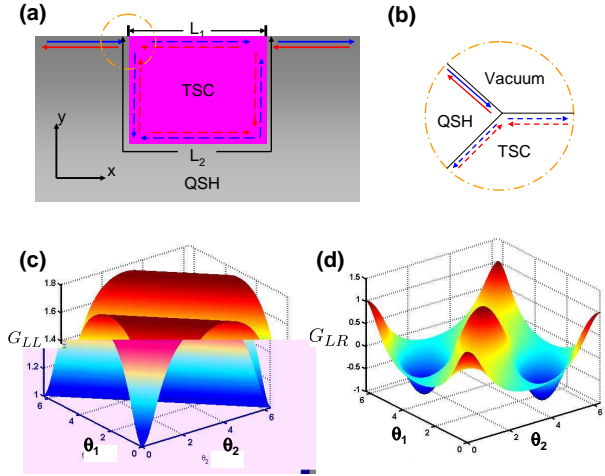}
    \end{center}
    \caption{ (Color online) (a) Proposed experimental setup for the HM interferometer; (b) Zoom into the Y junction, where Dirac fermions (full lines) are converted into Majorana fermions (dashed lines) and vice versa; (c) Dependence of the local conductance $G_{LL}=dI_L/dV_L$ and (d) the non-local conductance $G_{LR}=dI_L/dV_R$ on the phase factors $\theta_{1/2}$ (in units of $\frac{e^2}{h}$).}
    \label{fig:interferometer}
\end{figure}

In order to confirm the HM mode experimentally, we consider the setup shown in Fig.~\ref{fig:interferometer}(a). It is well-known that for a TI film, it is possible to obtain the two-dimensional quantum spin Hall state
with helical edge states by tuning the thickness of the film due to quantum confinement\cite{liu2010,lu2010}.
Then, for the TI film with proper thickness, we introduce an s-wave superconductor on both the top and bottom surfaces
near the upper edge as shown by the pink regime in Fig.~\ref{fig:interferometer}(a) (marked by the label TSC).
By tuning the relative phase of the pairing functions
for the SCs on the top and bottom surfaces, we can obtain a TSC with
HM modes at the edges. Hence, at the top edge in Fig.~\ref{fig:interferometer}(a), we find one helical Dirac fermion from the left hand side (LHS) will be splitted into two Majorana fermions and then recombined into another helical Dirac fermion at the right hand side (RHS). Therefore, a HM interferometer is obtained.

Next, we will discuss the transport properties of the HM interferometer within the scattering matrix formalism.
The HM interferometer can be divided into three parts, a Y junction at the LHS is connected to another Y junction
at the RHS by two freely propagating Majorana wires of different length $L_1$ and $L_2$.
The Y junction regime consists of one helical Dirac fermion mode and two Majorana fermion
modes, as shown in Fig.~\ref{fig:interferometer}(b). We name the Dirac fermion mode $|\psi^\alpha_{i(o)}\rangle$ ($\alpha=e,h$) and the two Majorana modes $|\gamma^\alpha_{i(o)}\rangle$ ($\alpha=1,2$), where i (o) stands for incoming (outgoing) mode. Time reversal requires
$T|\psi^\alpha_{i(o)}\rangle=(-)|\psi^\alpha_{o(i)}\rangle$ and $T|\gamma^\alpha_{i(o)}\rangle=(-)|\gamma^\alpha_{o(i)}\rangle$,
while particle-hole symmetry gives $C|\gamma^\alpha_{\eta}\rangle=|\gamma^\alpha_{\eta}\rangle$ and $C|\psi^{e(h)}_\eta\rangle
=|\psi^{h(e)}_\eta\rangle$ ($\eta=i,o$). The scattering wave function takes the form
$|\Psi\rangle=\sum_{\alpha,\eta}\left( c_{\alpha,\eta}|\gamma^\alpha_\eta\rangle+d_{\alpha,\eta}|\psi^\alpha_\eta \rangle\right)$,
and consequently the S-matrix for the Y junction $S_Y$ is defined as
\begin{eqnarray}
	\left(
	\begin{array}{c}
		c_{1,j,o}\\c_{2,j,o}\\d_{e,j,o}\\d_{h,j,o}
	\end{array}
	\right)=S_Y\left(
	\begin{array}{c}
		c_{1,j,i}\\c_{2,j,i}\\d_{e,j,i}\\d_{h,j,i}
	\end{array}
	\right),
	\label{eq:Smatdef}
\end{eqnarray}
where $j=L,R$ for the left and right Y junction.
With the constraints from time reversal symmetry\cite{bardarson2008}, particle-hole symmetry,
and the unitary condition, $S_Y$ can be parametrized as
\begin{eqnarray}
	S_Y=\left(
	\begin{array}{cc}
		\mathcal{R}_1&\mathcal{T}_1\\
		-\mathcal{T}_1^T&-e^{-i\phi}\mathcal{R}_1^*
	\end{array}
	\right),
	\label{eq:SmatY}
\end{eqnarray}
where
\begin{eqnarray}
	\mathcal{R}_1=\left(
	\begin{array}{cc}
		0&r_1\\
		-r_1&0\\
	\end{array}
	\right),
	\mathcal{T}_1=\left(
	\begin{array}{cc}
		t_1&t_2\\
		it^*_2&-it^*_1
	\end{array}
	\right),
	\label{eq:TR}
\end{eqnarray}
$|t_1|^2+|t_2|^2+|r_1|^2=1$, $r_1$ is real and $t_2=t^*_1$. In the latter equations, $r_1$ refers to the reflection amplitude, and $t_{1(2)}$ are the transmission amplitudes depending on the microscopic details.
In the middle regime, the two HM
fermions are propagating freely, only picking up a phase shift, therefore the corresponding
S-matrix $S_m$ reads
\begin{eqnarray}
	\left(
	\begin{array}{c}
		c_{1L,i}\\c_{2L,i}\\c_{1R,i}\\c_{2R,i}
	\end{array}
	\right)=S_m\left(
	\begin{array}{c}
		c_{1L,o}\\c_{2L,o}\\c_{1R,o}\\c_{2R,o}
	\end{array}
	\right),
	\label{eq:SmatdefTI}
\end{eqnarray}
with
\begin{eqnarray}
	&&S_m=\left(
	\begin{array}{cc}
		0&\mathcal{T}_{LR}\\
		\mathcal{T}_{RL}&0
	\end{array}
	\right),
	\mathcal{T}_{LR}=\left(
	\begin{array}{cc}
		e^{i\theta_{1}}&0\\
		0&e^{i\theta_{2}}\\
	\end{array}
	\right),
        \label{eq:Smatn2}
\end{eqnarray}
and $\mathcal{T}_{RL}=-\mathcal{T}_{LR}^T$ due to time reversal symmetry. The angles
$\theta_1$ and $\theta_2$ denote the phase shifts of the Majorana fermions
propagating along the two arms of the TSC. The phases (in the absence of trapped vortices in the TSC region) are given by $\theta_1 =k_x L_1 + \pi$, $\theta_2 =k_x L_2$,
where $L_{1,2}$ is the length of the upper (index 1) and the lower (index 2) arm in Fig.~\ref{fig:interferometer}(a). The $\pi$ phase shift, which we choose to add to $\theta_1$, comes from a Berry phase of a spin-1/2 rotation.
Combining the above S-matrix for three different regions, we can obtain the
total S-matrix $S_t$ defined as
\begin{eqnarray}
	\left(
	\begin{array}{c}
		d_{eL,o}\\d_{hL,o}\\d_{eR,o}\\d_{hR,o}
	\end{array}
	\right)=S_t\left(
	\begin{array}{c}
		d_{eL,i}\\d_{hL,i}\\d_{eR,i}\\d_{hR,i}
	\end{array}
	\right),
	\label{eq:Smatf2}
\end{eqnarray}
which connects the helical Dirac fermions on the LHS to those
on the RHS. The matrix elements of $S_t$ are named $s^{\alpha\beta}_{jl}$ where $\alpha,\beta=e,h$ denote electron or hole
and $j,l=L,R$ denote left or right lead. Then, the conductance is given by\cite{blonder1982,anantram1996,chung2010}
\begin{eqnarray}
	&&G_{jl}=\frac{d\langle \hat{I}_j\rangle}{d V_l}=\frac{e^2}{h}\left[ \delta_{jl}-T^{ee}_{jl}+T^{he}_{jl}  \right]
	\label{eq:Conductance1}
\end{eqnarray}
where $T^{\alpha\beta}_{jl}=|s^{\alpha\beta}_{jl}|^2$. $G_{jl}$ represents the differential (i.e. non-linear) conductance for the current in lead $j$ with respect to the voltage in lead $l$. We only consider the zero temperature case for simplicity.
Here, $T^{ee}_{jj}$ and $T^{ee}_{jl}$ ($j\neq l$) are the electron reflection and transmission probabilities, respectively, while
$T^{eh}_{jj}$ and $T^{eh}_{jl}$ ($j\neq l$) are the Andreev reflection and transmission probabilities, respectively.
By solving $S_t$ from the expressions (\ref{eq:Smatdef})-(\ref{eq:Smatn2}), we can obtain the $T^{\alpha\beta}_{jl}$. We find $T^{ee}_{LL}$ and $T^{ee}_{RR}$ vanish because of time reversal symmetry.
The Andreev reflection, the electron transmission, and the
Andreev transmission probabilities are given by
\begin{eqnarray}
	T^{he}_{LL}=T^{he}_{RR}=\frac{4r_1^2\sin^2\left( \frac{\theta_1+\theta_2}{2} \right)}{r_1^4+1-2r^2_1\cos(\theta_1+\theta_2)},
	\label{eq:TheLLRR}\\
	T^{ee}_{LR}=T^{ee}_{RL}
	=\frac{4|t_1|^4\sin^2(\frac{\theta_1-\theta_2}{2})}{r_1^4+1-2r_1^2\cos(\theta_1+\theta_2)},
	\label{eq:TeeLRRL}\\
	T^{he}_{LR}=T^{he}_{RL}=\frac{4|t_1|^4\cos^2(\frac{\theta_1-\theta_2}{2}) }
	{r_1^4+1-2r_1^2\cos(\theta_1+\theta_2)},
	\label{eq:TheLRRL}
\end{eqnarray}
respectively, which satisfy the current conservation condition $\sum_{l\beta}T^{\alpha\beta}_{jl}=1$.

From the expressions (\ref{eq:TheLLRR})-(\ref{eq:TheLRRL}), we find that the current through our HM interferometer
is qualitatively different from both (i) the current through a TI-SC-TI junction\cite{adroguer2010} and (ii) the current through a chiral Majorana interferometer\cite{fu2009,akhmerov2009}. First, although there is no electron reflection in the HM interferometer, Andreev reflection still exists. This is similar to the TI-SC-TI junction\cite{adroguer2010} but different to the chiral Majorana interferometer\cite{fu2009,akhmerov2009}. Second, in our case, an electron can transmit both as an electron or a hole, which can be used to distinguish the HM interferometer from the TI-SC-TI junction\cite{adroguer2010}, where only electron transmission is allowed.

In the HM interferometer, the normal and Andreev transmission probability depend on the phase difference between the upper and lower arms. It can be tuned by changing the applied bias, by varying the lengths of the arms $L_{1/2}$, or by varying the number of vortices trapped in the TSC regime as recently suggested in Ref.~\onlinecite{beri2011}.
The dependences of the local conductance $G_{LL}$ and the non-local conductance $G_{LR}$ on the parameters $\theta_1$ and $\theta_2$
are shown in Fig.~\ref{fig:interferometer}(c) and (d), respectively. The observation of such an interference pattern would be a clear signature of HM modes in this setup.

To summarize, we have proposed a realistic setup for a HM interferometer which can be used
to confirm the existence of HM modes and helical TSCs. The core parts of our setup could, for instance,
be realized in Bi$_2$Se$_3$ thin films selectively doped by copper to turn layers within the structure superconducting. 
A big advantage of our proposal is that the superconducting material and the TI material
are of the same type. Therefore, there should be no Schottky barriers between them maximizing the proximity effect. We have proposed to identify HM modes by measuring phase-dependent transport through a HM interferometer.

We would like to thank the Humboldt foundation (CXL), and the DFG-JST Research Unit ``Topological electronics''
(BT) for funding as well as B. Beri, M. Guigou, X.L. Qi, P. Recher, M. Wimmer, and S.-C. Zhang for interesting discussions.


\bibliography{helical}

\begin{thebibliography}{27}
\expandafter\ifx\csname natexlab\endcsname\relax\def\natexlab#1{#1}\fi
\expandafter\ifx\csname bibnamefont\endcsname\relax
  \def\bibnamefont#1{#1}\fi
\expandafter\ifx\csname bibfnamefont\endcsname\relax
  \def\bibfnamefont#1{#1}\fi
\expandafter\ifx\csname citenamefont\endcsname\relax
  \def\citenamefont#1{#1}\fi
\expandafter\ifx\csname url\endcsname\relax
  \def\url#1{\texttt{#1}}\fi
\expandafter\ifx\csname urlprefix\endcsname\relax\def\urlprefix{URL }\fi
\providecommand{\bibinfo}[2]{#2}
\providecommand{\eprint}[2][]{\url{#2}}

\bibitem[{\citenamefont{Hasan and Kane}(2010)}]{hasan2010}
\bibinfo{author}{\bibfnamefont{M.~Z.} \bibnamefont{Hasan}} \bibnamefont{and}
  \bibinfo{author}{\bibfnamefont{C.~L.} \bibnamefont{Kane}},
  \bibinfo{journal}{Rev. Mod. Phys.} \textbf{\bibinfo{volume}{82}},
  \bibinfo{pages}{3045} (\bibinfo{year}{2010}).

\bibitem[{\citenamefont{Qi and Zhang}(2010{\natexlab{a}})}]{qi2010}
\bibinfo{author}{\bibfnamefont{X.}~\bibnamefont{Qi}} \bibnamefont{and}
  \bibinfo{author}{\bibfnamefont{S.}~\bibnamefont{Zhang}},
  \bibinfo{journal}{Physics Today} \textbf{\bibinfo{volume}{63}},
  \bibinfo{pages}{33} (\bibinfo{year}{2010}{\natexlab{a}}).

\bibitem[{\citenamefont{Qi and Zhang}(2010{\natexlab{b}})}]{qi2010a}
\bibinfo{author}{\bibfnamefont{X.-L.} \bibnamefont{Qi}} \bibnamefont{and}
  \bibinfo{author}{\bibfnamefont{S.-C.} \bibnamefont{Zhang}},
  \bibinfo{howpublished}{e-print arXiv:1008.2026}
  (\bibinfo{year}{2010}{\natexlab{b}}).

\bibitem[{\citenamefont{Moore}(2010)}]{moore2010}
\bibinfo{author}{\bibfnamefont{J.~E.} \bibnamefont{Moore}},
  \bibinfo{journal}{Nature} \textbf{\bibinfo{volume}{464}},
  \bibinfo{pages}{194} (\bibinfo{year}{2010}).

\bibitem[{\citenamefont{Read and Green}(2000)}]{read2000}
\bibinfo{author}{\bibfnamefont{N.}~\bibnamefont{Read}} \bibnamefont{and}
  \bibinfo{author}{\bibfnamefont{D.}~\bibnamefont{Green}},
  \bibinfo{journal}{Phys. Rev. B} \textbf{\bibinfo{volume}{61}},
  \bibinfo{pages}{10267} (\bibinfo{year}{2000}).

\bibitem[{\citenamefont{Qi et~al.}(2009)\citenamefont{Qi, Hughes, Raghu, and
  Zhang}}]{qi2009b}
\bibinfo{author}{\bibfnamefont{X.-L.} \bibnamefont{Qi}},
  \bibinfo{author}{\bibfnamefont{T.~L.} \bibnamefont{Hughes}},
  \bibinfo{author}{\bibfnamefont{S.}~\bibnamefont{Raghu}}, \bibnamefont{and}
  \bibinfo{author}{\bibfnamefont{S.-C.} \bibnamefont{Zhang}},
  \bibinfo{journal}{Phys. Rev. Lett.} \textbf{\bibinfo{volume}{102}},
  \bibinfo{pages}{187001} (\bibinfo{year}{2009}).

\bibitem[{\citenamefont{Tanaka et~al.}(2009)\citenamefont{Tanaka, Yokoyama,
  Balatsky, and Nagaosa}}]{tanaka2009}
\bibinfo{author}{\bibfnamefont{Y.}~\bibnamefont{Tanaka}},
  \bibinfo{author}{\bibfnamefont{T.}~\bibnamefont{Yokoyama}},
  \bibinfo{author}{\bibfnamefont{A.~V.} \bibnamefont{Balatsky}},
  \bibnamefont{and} \bibinfo{author}{\bibfnamefont{N.}~\bibnamefont{Nagaosa}},
  \bibinfo{journal}{Phys. Rev. B} \textbf{\bibinfo{volume}{79}},
  \bibinfo{pages}{060505} (\bibinfo{year}{2009}).

\bibitem[{\citenamefont{Sato and Fujimoto}(2009)}]{sato2009}
\bibinfo{author}{\bibfnamefont{M.}~\bibnamefont{Sato}} \bibnamefont{and}
  \bibinfo{author}{\bibfnamefont{S.}~\bibnamefont{Fujimoto}},
  \bibinfo{journal}{Phys. Rev. B} \textbf{\bibinfo{volume}{79}},
  \bibinfo{pages}{094504} (\bibinfo{year}{2009}).

\bibitem[{\citenamefont{Ivanov}(2001)}]{ivanov2001}
\bibinfo{author}{\bibfnamefont{D.~A.} \bibnamefont{Ivanov}},
  \bibinfo{journal}{Phys. Rev. Lett.} \textbf{\bibinfo{volume}{86}},
  \bibinfo{pages}{268} (\bibinfo{year}{2001}).

\bibitem[{\citenamefont{Nayak et~al.}(2008)\citenamefont{Nayak, Simon, Stern,
  Freedman, and Das~Sarma}}]{nayak2008}
\bibinfo{author}{\bibfnamefont{C.}~\bibnamefont{Nayak}},
  \bibinfo{author}{\bibfnamefont{S.~H.} \bibnamefont{Simon}},
  \bibinfo{author}{\bibfnamefont{A.}~\bibnamefont{Stern}},
  \bibinfo{author}{\bibfnamefont{M.}~\bibnamefont{Freedman}}, \bibnamefont{and}
  \bibinfo{author}{\bibfnamefont{S.}~\bibnamefont{Das~Sarma}},
  \bibinfo{journal}{Rev. Mod. Phys.} \textbf{\bibinfo{volume}{80}},
  \bibinfo{pages}{1083} (\bibinfo{year}{2008}).

\bibitem[{\citenamefont{Fu and Kane}(2009)}]{fu2009}
\bibinfo{author}{\bibfnamefont{L.}~\bibnamefont{Fu}} \bibnamefont{and}
  \bibinfo{author}{\bibfnamefont{C.~L.} \bibnamefont{Kane}},
  \bibinfo{journal}{Phys. Rev. Lett.} \textbf{\bibinfo{volume}{102}},
  \bibinfo{pages}{216403} (\bibinfo{year}{2009}).

\bibitem[{\citenamefont{Akhmerov et~al.}(2009)\citenamefont{Akhmerov, Nilsson,
  and Beenakker}}]{akhmerov2009}
\bibinfo{author}{\bibfnamefont{A.~R.} \bibnamefont{Akhmerov}},
  \bibinfo{author}{\bibfnamefont{J.}~\bibnamefont{Nilsson}}, \bibnamefont{and}
  \bibinfo{author}{\bibfnamefont{C.~W.~J.} \bibnamefont{Beenakker}},
  \bibinfo{journal}{Phys. Rev. Lett.} \textbf{\bibinfo{volume}{102}},
  \bibinfo{pages}{216404} (\bibinfo{year}{2009}).

\bibitem[{foo()}]{footnote1}
\bibinfo{note}{Recently, a similar detection scheme for HM modes as the one
  discussed by us in this paper has been independently proposed by
  Beri\cite{beri2011}. However, in our work, we additionally propose a new type
  of setup in which the HM interferometer can actually be realized based on
  existing materials.}

\bibitem[{\citenamefont{Wray et~al.}(2010)\citenamefont{Wray, Xu, Xia, Hor,
  Qian, Fedorov, Lin, Bansil, Cava, and Hasan}}]{wray2010}
\bibinfo{author}{\bibfnamefont{L.~A.} \bibnamefont{Wray}},
  \bibinfo{author}{\bibfnamefont{S.}~\bibnamefont{Xu}},
  \bibinfo{author}{\bibfnamefont{Y.}~\bibnamefont{Xia}},
  \bibinfo{author}{\bibfnamefont{Y.~S.} \bibnamefont{Hor}},
  \bibinfo{author}{\bibfnamefont{D.}~\bibnamefont{Qian}},
  \bibinfo{author}{\bibfnamefont{A.~V.} \bibnamefont{Fedorov}},
  \bibinfo{author}{\bibfnamefont{H.}~\bibnamefont{Lin}},
  \bibinfo{author}{\bibfnamefont{A.}~\bibnamefont{Bansil}},
  \bibinfo{author}{\bibfnamefont{R.~J.} \bibnamefont{Cava}}, \bibnamefont{and}
  \bibinfo{author}{\bibfnamefont{M.~Z.} \bibnamefont{Hasan}},
  \bibinfo{journal}{Nature Phys.} \textbf{\bibinfo{volume}{6}},
  \bibinfo{pages}{855} (\bibinfo{year}{2010}).

\bibitem[{\citenamefont{Yan et~al.}(2010)\citenamefont{Yan, Liu, Zhang, Yam,
  Qi, Frauenheim, and Zhang}}]{yan2010}
\bibinfo{author}{\bibfnamefont{B.}~\bibnamefont{Yan}},
  \bibinfo{author}{\bibfnamefont{C.}~\bibnamefont{Liu}},
  \bibinfo{author}{\bibfnamefont{H.}~\bibnamefont{Zhang}},
  \bibinfo{author}{\bibfnamefont{C.}~\bibnamefont{Yam}},
  \bibinfo{author}{\bibfnamefont{X.}~\bibnamefont{Qi}},
  \bibinfo{author}{\bibfnamefont{T.}~\bibnamefont{Frauenheim}},
  \bibnamefont{and} \bibinfo{author}{\bibfnamefont{S.}~\bibnamefont{Zhang}},
  \bibinfo{journal}{Europhys. Lett.} \textbf{\bibinfo{volume}{90}},
  \bibinfo{pages}{37002} (\bibinfo{year}{2010}).

\bibitem[{\citenamefont{Zhang et~al.}(2011)\citenamefont{Zhang, Zhang, Weng,
  Zhang, Yang, Liu, Feng, Wang, Yu, Cao et~al.}}]{zhang2011}
\bibinfo{author}{\bibfnamefont{J.~L.} \bibnamefont{Zhang}},
  \bibinfo{author}{\bibfnamefont{S.~J.} \bibnamefont{Zhang}},
  \bibinfo{author}{\bibfnamefont{H.~M.} \bibnamefont{Weng}},
  \bibinfo{author}{\bibfnamefont{W.}~\bibnamefont{Zhang}},
  \bibinfo{author}{\bibfnamefont{L.~X.} \bibnamefont{Yang}},
  \bibinfo{author}{\bibfnamefont{Q.~Q.} \bibnamefont{Liu}},
  \bibinfo{author}{\bibfnamefont{S.~M.} \bibnamefont{Feng}},
  \bibinfo{author}{\bibfnamefont{X.~C.} \bibnamefont{Wang}},
  \bibinfo{author}{\bibfnamefont{R.~C.} \bibnamefont{Yu}},
  \bibinfo{author}{\bibfnamefont{L.~Z.} \bibnamefont{Cao}},
  \bibnamefont{et~al.}, \bibinfo{journal}{Proceedings of the National Academy
  of Sciences} \textbf{\bibinfo{volume}{108}}, \bibinfo{pages}{24 }
  (\bibinfo{year}{2011}).

\bibitem[{\citenamefont{Liu et~al.}(2010)\citenamefont{Liu, Zhang, Yan, Qi,
  Frauenheim, Dai, Fang, and Zhang}}]{liu2010}
\bibinfo{author}{\bibfnamefont{C.}~\bibnamefont{Liu}},
  \bibinfo{author}{\bibfnamefont{H.}~\bibnamefont{Zhang}},
  \bibinfo{author}{\bibfnamefont{B.}~\bibnamefont{Yan}},
  \bibinfo{author}{\bibfnamefont{X.}~\bibnamefont{Qi}},
  \bibinfo{author}{\bibfnamefont{T.}~\bibnamefont{Frauenheim}},
  \bibinfo{author}{\bibfnamefont{X.}~\bibnamefont{Dai}},
  \bibinfo{author}{\bibfnamefont{Z.}~\bibnamefont{Fang}}, \bibnamefont{and}
  \bibinfo{author}{\bibfnamefont{S.}~\bibnamefont{Zhang}},
  \bibinfo{journal}{Phys. Rev. B} \textbf{\bibinfo{volume}{81}},
  \bibinfo{pages}{041307} (\bibinfo{year}{2010}).

\bibitem[{\citenamefont{Lu et~al.}(2010)\citenamefont{Lu, Shan, Yao, Niu, and
  Shen}}]{lu2010}
\bibinfo{author}{\bibfnamefont{H.}~\bibnamefont{Lu}},
  \bibinfo{author}{\bibfnamefont{W.}~\bibnamefont{Shan}},
  \bibinfo{author}{\bibfnamefont{W.}~\bibnamefont{Yao}},
  \bibinfo{author}{\bibfnamefont{Q.}~\bibnamefont{Niu}}, \bibnamefont{and}
  \bibinfo{author}{\bibfnamefont{S.}~\bibnamefont{Shen}},
  \bibinfo{journal}{Phys. Rev. B} \textbf{\bibinfo{volume}{81}},
  \bibinfo{pages}{115407} (\bibinfo{year}{2010}).

\bibitem[{\citenamefont{\textrm{B. A. Bernevig}
  et~al.}(2006)\citenamefont{\textrm{B. A. Bernevig}, \textrm{T. L. Hughes},
  and \textrm{S.C. Zhang}}}]{bernevig2006d}
\bibinfo{author}{\bibnamefont{\textrm{B. A. Bernevig}}},
  \bibinfo{author}{\bibnamefont{\textrm{T. L. Hughes}}}, \bibnamefont{and}
  \bibinfo{author}{\bibnamefont{\textrm{S.C. Zhang}}},
  \bibinfo{journal}{Science} \textbf{\bibinfo{volume}{314}},
  \bibinfo{pages}{1757} (\bibinfo{year}{2006}).

\bibitem[{\citenamefont{Qi et~al.}(2010)\citenamefont{Qi, Hughes, and
  Zhang}}]{qi2010c}
\bibinfo{author}{\bibfnamefont{X.-L.} \bibnamefont{Qi}},
  \bibinfo{author}{\bibfnamefont{T.~L.} \bibnamefont{Hughes}},
  \bibnamefont{and} \bibinfo{author}{\bibfnamefont{S.-C.} \bibnamefont{Zhang}},
  \bibinfo{howpublished}{e-print arXiv:1003.5448} (\bibinfo{year}{2010}).

\bibitem[{\citenamefont{Fu and Kane}(2008)}]{fu2008}
\bibinfo{author}{\bibfnamefont{L.}~\bibnamefont{Fu}} \bibnamefont{and}
  \bibinfo{author}{\bibfnamefont{C.~L.} \bibnamefont{Kane}},
  \bibinfo{journal}{Phys. Rev. Lett.} \textbf{\bibinfo{volume}{100}},
  \bibinfo{pages}{096407} (\bibinfo{year}{2008}).

\bibitem[{\citenamefont{Bardarson}(2008)}]{bardarson2008}
\bibinfo{author}{\bibfnamefont{J.~H.} \bibnamefont{Bardarson}},
  \bibinfo{journal}{Journal of Physics A: Mathematical and Theoretical}
  \textbf{\bibinfo{volume}{41}}, \bibinfo{pages}{405203}
  (\bibinfo{year}{2008}).

\bibitem[{\citenamefont{Blonder et~al.}(1982)\citenamefont{Blonder, Tinkham,
  and Klapwijk}}]{blonder1982}
\bibinfo{author}{\bibfnamefont{G.~E.} \bibnamefont{Blonder}},
  \bibinfo{author}{\bibfnamefont{M.}~\bibnamefont{Tinkham}}, \bibnamefont{and}
  \bibinfo{author}{\bibfnamefont{T.~M.} \bibnamefont{Klapwijk}},
  \bibinfo{journal}{Phys. Rev. B} \textbf{\bibinfo{volume}{25}},
  \bibinfo{pages}{4515} (\bibinfo{year}{1982}).

\bibitem[{\citenamefont{Anantram and Datta}(1996)}]{anantram1996}
\bibinfo{author}{\bibfnamefont{M.~P.} \bibnamefont{Anantram}} \bibnamefont{and}
  \bibinfo{author}{\bibfnamefont{S.}~\bibnamefont{Datta}},
  \bibinfo{journal}{Phys. Rev. B} \textbf{\bibinfo{volume}{53}},
  \bibinfo{pages}{16390} (\bibinfo{year}{1996}).

\bibitem[{\citenamefont{Chung et~al.}(2010)\citenamefont{Chung, Qi, Maciejko,
  and Zhang}}]{chung2010}
\bibinfo{author}{\bibfnamefont{S.~B.} \bibnamefont{Chung}},
  \bibinfo{author}{\bibfnamefont{X.}~\bibnamefont{Qi}},
  \bibinfo{author}{\bibfnamefont{J.}~\bibnamefont{Maciejko}}, \bibnamefont{and}
  \bibinfo{author}{\bibfnamefont{S.}~\bibnamefont{Zhang}},
  \bibinfo{howpublished}{e-print arXiv:1008.2003} (\bibinfo{year}{2010}).

\bibitem[{\citenamefont{Adroguer et~al.}(2010)\citenamefont{Adroguer, Grenier,
  Carpentier, Cayssol, Degiovanni, and Orignac}}]{adroguer2010}
\bibinfo{author}{\bibfnamefont{P.}~\bibnamefont{Adroguer}},
  \bibinfo{author}{\bibfnamefont{C.}~\bibnamefont{Grenier}},
  \bibinfo{author}{\bibfnamefont{D.}~\bibnamefont{Carpentier}},
  \bibinfo{author}{\bibfnamefont{J.}~\bibnamefont{Cayssol}},
  \bibinfo{author}{\bibfnamefont{P.}~\bibnamefont{Degiovanni}},
  \bibnamefont{and} \bibinfo{author}{\bibfnamefont{E.}~\bibnamefont{Orignac}},
  \bibinfo{journal}{Phys. Rev. B} \textbf{\bibinfo{volume}{82}},
  \bibinfo{pages}{081303} (\bibinfo{year}{2010}).

\bibitem[{\citenamefont{Beri}(2011)}]{beri2011}
\bibinfo{author}{\bibfnamefont{B.}~\bibnamefont{Beri}},
  \bibinfo{howpublished}{e-print arXiv:1102.4541} (\bibinfo{year}{2011}).

\end{thebibliography}

\end{document}